\documentstyle[12pt]{article}

  \def\pp{{\mathchoice
          {
              \kern 1pt%
              \raise 1pt
              \vbox{\hrule width5pt height0.4pt depth0pt
                    \kern -2pt
                    \hbox{\kern 2.3pt
                          \vrule width0.4pt height6pt depth0pt
                          }
                    \kern -2pt
                    \hrule width5pt height0.4pt depth0pt}%
                    \kern 1pt
           }
            {
              \kern 1pt%
              \raise 1pt
              \vbox{\hrule width4.3pt height0.4pt depth0pt
                    \kern -1.8pt
                    \hbox{\kern 1.95pt
                          \vrule width0.4pt height5.4pt depth0pt
                          }
                    \kern -1.8pt
                    \hrule width4.3pt height0.4pt depth0pt}%
                    \kern 1pt
            }
            {
              \kern 0.5pt%
              \raise 1pt
              \vbox{\hrule width4.0pt height0.3pt depth0pt
                    \kern -1.9pt  
                    \hbox{\kern 1.85pt
                          \vrule width0.3pt height5.7pt depth0pt
                          }
                    \kern -1.9pt
                    \hrule width4.0pt height0.3pt depth0pt}%
                    \kern 0.5pt
            }
            {
              \kern 0.5pt%
              \raise 1pt
              \vbox{\hrule width3.6pt height0.3pt depth0pt
                    \kern -1.5pt
                    \hbox{\kern 1.65pt
                          \vrule width0.3pt height4.5pt depth0pt
                          }
                    \kern -1.5pt
                    \hrule width3.6pt height0.3pt depth0pt}%
                    \kern 0.5pt
            }
        }}

  \def\mm{{\mathchoice
                       {
                             \kern 1pt
               \raise 1pt    \vbox{\hrule width5pt height0.4pt depth0pt
                                  \kern 2pt
                                  \hrule width5pt height0.4pt depth0pt}
                             \kern 1pt}
                       {
                            \kern 1pt
               \raise 1pt \vbox{\hrule width4.3pt height0.4pt depth0pt
                                  \kern 1.8pt
                                  \hrule width4.3pt height0.4pt depth0pt}
                             \kern 1pt}
                       {
                            \kern 0.5pt
               \raise 1pt
                            \vbox{\hrule width4.0pt height0.3pt depth0pt
                                  \kern 1.9pt
                                  \hrule width4.0pt height0.3pt depth0pt}
                            \kern 1pt}
                       {
                           \kern 0.5pt
             \raise 1pt  \vbox{\hrule width3.6pt height0.3pt depth0pt
                                  \kern 1.5pt
                                  \hrule width3.6pt height0.3pt depth0pt}
                           \kern 0.5pt}
                       }}

\catcode`@=11
\def\un#1{\relax\ifmmode\@@underline#1\else
        $\@@underline{\hbox{#1}}$\relax\fi}
\catcode`@=12

\let\du=\du                     

\def\a{\alpha}

\def\q{\theta}

\def\s{\sigma}
\def\t{\tau}

\def\bo{{\raise-.5ex\hbox{\large$\Box$}}}               
\def\pr{\prod}                                          
\def\TH{{\raise.2ex\hbox{$\displaystyle \bigodot$}\mskip-4.7mu \llap H \;}}
\def\face{{\raise.2ex\hbox{$\displaystyle \bigodot$}\mskip-2.2mu \llap {$\ddot
        \smile$}}}                                      

\def\sp#1{{}^{#1}}                              
\def\leftrightarrowfill{$\mathsurround=0pt \mathord\leftarrow \mkern-6mu
        \cleaders\hbox{$\mkern-2mu \mathord- \mkern-2mu$}\hfill
        \mkern-6mu \mathord\rightarrow$}
\def\dvec#1{\vbox{\ialign{##\crcr
        \leftrightarrowfill\crcr\noalign{\kern-1pt\nointerlineskip}
        $\hfil\displaystyle{#1}\hfil$\crcr}}}           

\def\frac#1#2{{\textstyle{#1\over\vphantom2\smash{\raise.20ex
        \hbox{$\scriptstyle{#2}$}}}}}                   
\def\sfrac#1#2{{\vphantom1\smash{\lower.5ex\hbox{\small$#1$}}\over
        \vphantom1\smash{\raise.4ex\hbox{\small$#2$}}}} 
\def\bfrac#1#2{{\vphantom1\smash{\lower.5ex\hbox{$#1$}}\over
        \vphantom1\smash{\raise.3ex\hbox{$#2$}}}}       
\def\afrac#1#2{{\vphantom1\smash{\lower.5ex\hbox{$#1$}}\over#2}}    

\def\[{\lfloor{\hskip 0.35pt}\!\!\!\lceil}
\def\]{\rfloor{\hskip 0.35pt}\!\!\!\rceil}

\def\du#1#2{_{#1}{}^{#2}}

\def\un{\underline}

\def\low#1{{\raise -3pt\hbox{${\hskip 0.75pt}\!_{#1}$}}}

\newskip\humongous \humongous=0pt plus 1000pt minus 1000pt

\newif\ifdtup

\def\plpl{\raise-2pt\hbox{$\raise3pt\hbox{$_+$}\hskip-6.67pt\raise0.0pt}}
\def\mimi{\raise-2pt\hbox{$\raise3pt\hbox{$_-$}\hskip-6.67pt\raise0.0pt}}

\def\dvm{\raisebox{-.145ex}{\rlap{$=$}}}
\def\DM{{\scriptsize{\dvm}}~~}

\def\ref#1{$\sp{#1)}$}

\def\pl#1#2#3{Phys.~Lett.~{\bf {#1}B} (19{#2}) #3}
\def\np#1#2#3{Nucl.~Phys.~{\bf B{#1}} (19{#2}) #3}
\def\prl#1#2#3{Phys.~Rev.~Lett.~{\bf #1} (19{#2}) #3}
\def\pr#1#2#3{Phys.~Rev.~{\bf D{#1}} (19{#2}) #3}
\def\cqg#1#2#3{Class.~and Quantum Grav.~{\bf {#1}} (19{#2}) #3}

\def\mpl#1#2#3{Mod.~Phys.~Lett.~{\bf A{#1}} (19{#2}) #3}

\def\ibid#1#2#3{{\it ibid.}~{\bf {#1}} (19{#2}) #3}

\parskip=\medskipamount                 
\thicklines                         

\begin{document}


\thispagestyle{empty}               

\def\border{                                            
        \setlength{\unitlength}{1mm}
        \newcount\xco
        \newcount\yco
        \xco=-24
        \yco=12
        \begin{picture}(140,0)
        \put(-20,11){\tiny Institut f\"ur Theoretische Physik Universit\"at
Hannover~~ Institut f\"ur Theoretische Physik Universit\"at Hannover~~
Institut f\"ur Theoretische Physik Hannover}
        \put(-20,-241.5){\tiny Institut f\"ur Theoretische Physik Universit\"at
Hannover~~ Institut f\"ur Theoretische Physik Universit\"at Hannover~~
Institut f\"ur Theoretische Physik Hannover}
        \end{picture}
        \par\vskip-8mm}

\def\headpic{                                           
        \indent
        \setlength{\unitlength}{.8mm}
        \thinlines
        \par
        \begin{picture}(29,16)
        \put(75,16){\line(1,0){4}}
        \put(80,16){\line(1,0){4}}
      \put(85,16){\line(1,0){4}}
        \put(92,16){\line(1,0){4}}

        \put(85,0){\line(1,0){4}}
        \put(89,8){\line(1,0){3}}
        \put(92,0){\line(1,0){4}}

        \put(85,0){\line(0,1){16}}
        \put(96,0){\line(0,1){16}}
        \put(92,16){\line(1,0){4}}

        \put(85,0){\line(1,0){4}}
        \put(89,8){\line(1,0){3}}
        \put(92,0){\line(1,0){4}}

        \put(85,0){\line(0,1){16}}
        \put(96,0){\line(0,1){16}}
        \put(79,0){\line(0,1){16}}
        \put(80,0){\line(0,1){16}}
        \put(89,0){\line(0,1){16}}
        \put(92,0){\line(0,1){16}}
        \put(79,16){\oval(8,32)[bl]}
        \put(80,16){\oval(8,32)[br]}

        \end{picture}
        \par\vskip-6.5mm
        \thicklines}

\border\headpic {\hbox to\hsize{
\vbox{\noindent ITP--UH--09/96 \hfill June 1996 \\
hep-th/9606142 \hfill }}}

\noindent
\vskip1.3cm
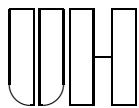
\begin{center}

{\Large\bf {}From N=2 Strings to F~\&~M Theory}~\footnote{Supported in part by 
the `Deutsche Forschungsgemeinschaft' and the `Volkswagen Stiftung'}
${}^{,}$~\footnote{Talk given at the International Workshop 
`{\it Integrable Models and Strings},
24--25 June, 1996, \newline ${~~~~~}$ Garbsen, Germany}\\

\vglue.3in

Sergei V. Ketov \footnote{
On leave of absence from:
High Current Electronics Institute of the Russian Academy of Sciences,
\newline ${~~~~~}$ Siberian Branch, Akademichesky~4, Tomsk 634055, Russia}

{\it Institut f\"ur Theoretische Physik, Universit\"at Hannover}\\
{\it Appelstra\ss{}e 2, 30167 Hannover, Germany}\\
{\sl ketov@itp.uni-hannover.de}
\end{center}
\vglue.2in
\begin{center}
{\Large\bf Abstract}
\end{center}

\noindent
Taking the N=2 strings as the starting point, we discuss the equivalent
self-dual field theories and analyse their symmetry structure in $2+2$ 
dimensions. Restoring the full `Lorentz' invariance in the target space 
necessarily leads to an extension of the N=2 string theory to a theory of 
$2+2$ dimensional {\it supermembranes} propagating in $2+10$ dimensional 
target space. The supermembrane requires {\it maximal} conformal supersymmetry
in $2+2$ dimensions, in the way advocated by Siegel. The corresponding 
self-dual N=4 Yang-Mills theory and the
self-dual N=8 (gauged) supergravity in 2+2 dimensions thus appear to be 
naturally associated to the membrane theory, not a string. Since the same 
theory of membranes seems to represent the M-theory which is apparently 
underlying the all known N=1 string theories, the N=2 strings now appear on 
equal footing with the other string models as particular limits of the 
unique fundamental theory. Unlike the standard $10$-dimensional superstrings, 
the N=2 strings seem to be much closer to a membrane description of the 
F~\&~M theory.

\newpage

\section{Introduction}

Since the discovery of string dualities, much evidence was collected in favor
that various `different' string theories can be understood as particular 
limits of a unique underlying theory whose basic formulation is yet to be 
found. It seems also that the fundamental theory is not just a theory of 
strings but it describes fields, strings and membranes in a unified way. There
is a candidate for such unified theory -- the so-called 
{\it M-theory}~\cite{sch1,w1}~\footnote{See ref.~\cite{hull} for a review.} or
its refined {\it F-theory} formulation~\cite{va1}-- which can be reduced to 
all known 10-dimensional supertrings and 11-dimensional supergravity as well. 
Accordingly, there should be a similar way to understand strings with the 
{\it extended} world-sheet supersymmetry -- the so-called N=2 strings -- in
terms of the M-theory. 

As was noticed recently~\cite{kmo}, the $N=(2,1)$ heterotic
string is not only connected to the M-theory in particular backgrounds, but it
also suggests the M-theory definition as a theory of $2+2$ dimensional 
membranes (sometimes called {\it M-branes}~\cite{town1}) embedded in $2+10$ 
dimensions with a null reduction. If so, the origin of M-branes should be
understood from the basic properties of N=2 strings. It is the purpose of this
paper to argue that the hidden world-volume and the membrane target space 
dimensions are in fact {\it required} by natural symmetries which are broken 
in the known N=2 string formulations. By the natural symmetries I mean the 
`Lorentz' invariance and supersymmetry in $2+2$ dimensions. These symmetries 
also uniquely determine the dymanics of M-branes which is given by the 
self-dual gauged supergravity with the maximally extended N=8 supersymmetry. 
The relevant maximally extended self-dual field theories were constructed some 
time ago by Siegel~\cite{sie} in the light-cone gauge (see also 
ref.~\cite{kgn}) but, unlike the earlier expectations, they appear not to be
related to the N=2 strings, but to the M-branes. The suggestion that the 
M-branes should be described by a kind of self-dual gravity coupled to  a
self-dual matter also appeared in ref.~\cite{kmo}. It is the goal of this 
paper to specify  the symmetries of this self-dual field theory. Unlike the
analysis of the $1+1$ and $1+2$ dimensional target space versions of the 
$N=(2,1)$ strings in ref.~\cite{kmo}, we impose the $2+2$ dimensional 
`Lorentz' symmetry as the crucial symmetry of M-branes.
\vglue.2in

\section{N=2 string symmetries}

The N=2 strings are strings with two world-sheet (local) supersymmetries. There
exist $N=(2,2)$ open and closed strings, and $N=(2,1)$ and $N=(2,0)$ heterotic 
strings. The N=2 strings have twenty years-long history.~\footnote{See 
ref.~\cite{ade} for the first references on the subject, and 
refs.~\cite{markus,ke1,book} for a review.} The gauge-invariant $N=2$ string 
world-sheet actions in the NSR-type formulation are given by couplings of a 
two-dimensional N=2 
supergravity to a complex N=2 scalar matter~\cite{bs}. Gauge-fixing produces
conformal ghosts, complex superconformal ghosts and real abelian ghosts, as
usual. The corresponding (chiral) world-sheet current algebras include a 
stress-tensor, two supercurrents and an abelian current; taken together, they 
constitute an N=2 superconformal algebra. As a result, the critical closed and
open N=2 strings live in four dimensions with the signature $2+2$.~\footnote{
The signature is dictated by the existence of a complex structure and 
non-trivial kinematics.} The current algebras of the N=2 heterotic strings 
include an additional abelian null current needed for the nilpotency of the 
BRST charge, and it effectively reduces the target spacetime dynamics down to 
$1+2$ or $1+1$ dimensions~\cite{ov}. The full target space dimension (with the 
heterotic modes) is $2+26$ for the left-moving modes of the $N=(2,0)$ 
heterotic string and $2+10$ for the $N=(2,1)$ ones, respectively.

The BRST cohomology and on-shell amplitudes of N=2 strings were 
investigated in detail by several groups~\cite{ov,bov,ha,te}. The results of 
that investigations can be summarized as follows. There exist only one 
massless particle in the open or closed N=2 string spectrum. This particle can
be identified with the Yang 
scalar of self-dual Yang-Mills theory for open strings, or the K\"ahler scalar
of self-dual supergravity for closed strings, while infinitely many massive
string modes are all unphysical. The NS- and R-type states are connected by 
the spectral flow, which is also a symmetry of correlation 
functions. Accordingly, the N=2 strings lack `space-time' supersymmetry.
Twisting the N=2 superconformal algebra yields some additional twisted 
physical states which would-be the target space `fermions', but they actually 
decouple. The corresponding `space-time fermionic' vertex operators 
anticommute modulo picture-changing, instead of producing `space-time' 
translations, as required by `space-time' supersymmetry~\cite{ha}. The only 
non-vanishing scattering amplitudes appear to be 3-point trees (and, maybe, 
3-point loops as well), while all the 
other tree and loop N=2 string amplitudes apparently vanish due to kinematical
reasons. As a result, an N=2 string theory appears to be equivalent to a 
self-dual field theory. In particular, the N=2 open string amplitudes are 
reproduced by either the Yang non-linear sigma-model action~\cite{yang} or the
Leznov-Parkes cubic action~\cite{lp}, following from a field integration of 
the self-dual Yang-Mills equations in particular Lorentz non-covariant gauges,
 and related to each other by a duality transformation. As far as the closed 
N=2 strings are concerned, the equivalent non-covariant field theory action is
 known as the Plebanski action~\cite{ple} for the self-dual gravity. 

One generically finds more massless physical states in the heterotic 
N=2 string spectra. In particular, the $(2,1)$ heterotic string has 8 bosonic 
and 8 fermionic massless particles in $1+2$ dimensions. The equivalent field 
theory is given by a three-dimensional coupling of self-dual Yang-Mills and 
self-dual gravity~\cite{ov}. Unlike the N=0 and N=1 strings, the N=2 string
world-sheet symmetries do not allow massive string excitations to be physical. 

The natural global continuous (`Lorentz') symmetry of a flat $2+2$ dimensional
target space (`space-time') is $SO(2,2)\cong SU(1,1) \otimes SU(1,1)\cong
SL(2,{\bf R})\otimes SL(2,{\bf R})$. The NSR-type N=2 string actions used to
calculate the N=2 string amplitudes have only a part of it, namely, 
$U(1,1)\cong SU(1,1)\otimes U(1)$ or $GL(2,{\bf R})$, so is the symmetry of
amplitudes in the absence of world-sheet abelian instantons.~\footnote{Taking
into account N=2 string amplitudes with a non-trivial Chern class (or $U(1)$
instanton \newline ${~~~~~}$ number) makes the symmetry even lower~\cite{ha}.}

Adding to the N=2 string generators of the N=2 superconformal algebra the 
spectral flow operator and its inverse provides the raising and lowering 
operators of $SU(1,1)$. Closing the algebra, one extends the initial N=2 
superconformal algebra to the `small' twisted N=4 superconformal algebra. This
remarkable property allows one to treat the N=2 string theory as an N=4 
topological field theory~\cite{bov}.~\footnote{It does not, however, mean that
the critical ~N=4~ strings are `the same' as the critical ~N=2 
\newline ${~~~~~}$ strings.} It is even more important that this 
reformulation brings the additional internal symmetry, $SU(1,1)$, which is just
needed to restore the broken `Lorentz' symmetry $U(1,1)$ to the $SO(2,2)$. The
embeddings of the N=2 algebra into the N=4 one are parameterized by vielbeins 
(twistors) belonging to the harmonic space $SU(1,1)/U(1)$, which is the space
of all complex structures in $2+2$ `space-time'. The harmonic space 
technically adds two additional world-sheet dimensions to an N=2 string. It 
now becomes obvious that, in order to get back the `Lorentz' symmetry in the 
N=2 string theory target space, one has to take the two additional harmonic 
(one time-like and another space-like) dimensions for real, by {\it extending}
the N=2 string world-sheet to the $2+2$ dimensional 
world-volume (M-brane), i.e. to complexify the N=2 string world-sheet 
coordinates $\t$ and $\s$.~\footnote{The idea of the string world-sheet 
complexification also appeared in the stidies of high-energy \newline
${~~~~~}$ behavior of string theories~\cite{wi2}.}

The target space for M-branes, where they are supposed to propagate, can
also be fixed by merely restoring the `space-time Lorentz' symmetry of the N=2
strings, as we are now going to argue.
 
\section{~$2+10$~ out of ~$2+2$~}

It was first noticed by Siegel~\cite{sie} that the self-duality and `Lorentz' 
invariance in $2+2$ dimensions imply the {\it maximal} supersymmetry. 
It becomes transparent in the light-cone gauge for self-dual field theories,
where only physical degrees of freedom are kept.

The irreducible massless representations of $SO(2,2)\cong SL(2,{\bf R})'\otimes
SL(2,{\bf R})$ are either self-dual or chiral, and they are all real and 
one-dimensional. It is therefore convenient to introduce the basis, in which
a $2+2$ dimensional vector has components $x^{\a',\a}$, so that 
the first helicity index $\a'=(+',-')$ refers to the first $SL(2,{\bf R})'$
component while $\a=(+,-)$ to the second one. In this basis, $x^{\a',-}$ are
treated as `time' coordinates, whereas $x^{\a',+}$ as `space' coordinates. In 
the light-cone gauge, a self-dual gauge theory is described in terms of a 
single field -- the so-called {\it prepotential} $V(x)$ -- whose helicity is 
$+1$ for the self-dual Yang-Mills theory $(V_{\DM})$, and is $+2$ for the 
self-dual gravity $(V_{\DM\DM})$. Note that going to the light-cone gauge 
already breaks the `Lorentz' invariance down to $SL(2,{\bf R})'\otimes GL(1)$.
The $N$-supersymmetrization of the light-cone gauge description of a self-dual 
gauge theory is straightforward: one should simply extend the prepotential to
an $N$-extended real chiral superfield $V(x,\q^+)$ to be also dependent on $N$ 
anticommuting (Grassmannian) self-dual superspace coordinates $\q^{A,+}$,
$A=1,2,\ldots,N$, which are Majorana-Weyl spinors in $2+2$ dimensions.

A free field theory action in the light-cone gauge takes the universal form,
$$ I_{\rm free} =\frac{1}{2}
\int d^{2+2}xd^N\q^+\,V_{-,\ldots}\bo V_{-,\ldots}~, \eqno(1)$$
and it simultaneously determines the field content of the theory. For generic 
$N$, the $I_{\rm free}$ is not `Lorentz'-invariant, but it becomes invariant
when the light-cone superfield $V$ is self-dual or, equivalently, when the
$GL(1)$ charge of the superspace measure cancels that of the integrand in
eq.~(1). Both requirements obviously imply $N=N_{\rm max}$. For example, 
the first component of the maximally extended N=4 supersymmetric self-dual 
Yang-Mills prepotential $V_{\DM}$ has the helicity $+1$, whereas its last 
component has helicity $-1$, which is just needed for a `Lorentz'-invariant 
action. Similarly, one finds that the free action (1) in terms of the 
$N$-extended self-dual supergravity prepotential $V_{\DM\DM}$ requires the 
$N=8$ supersymmetry in order to be `Lorentz'-invariant. Siegel gave the 
light-cone formulations for the maximally supersymmetric self-dual gauge and
gravity field theories, both in components and in self-dual 
superspace~\cite{sie}. The full (interacting) theories he constructed are 
actually quite similar to the non-self-dual supersymmetric gauge theories to
be formulated in the light-cone gauge~\cite{lc3}.

As far as the heterotic N=2 `strings' are concerned, in the $N=(2,0)$ case
one gets the N=4 supersymmetric coupling of self-dual super-Yang-Mills to 
self-dual supergravity. In the $N=(2,1)$ case,~\footnote{It was 
suggested~\cite{kmo} that the $N=(2,1)$ heterotic strings describe the strong 
coupling dynamics \newline ${~~~~~}$ of the ten-dimensional superstrings 
compactified down to two dimensions.}  the `Lorentz' 
invariance still requires N=8, while the gauge group for the heterotic vector 
bosons is obviously restricted to $SO(8)$ or its non-compact version. 

The appearance of the $SO(8)$ internal symmetry in the maximally supersymmetric
heterotic case is quite remarkable. Having substituted N=2 strings by 
M-branes, we thus restored not only the `Lorentz' $SO(2,2)$ symmetry but the 
$N=8$ supersymmetry and the $SO(8)$ internal symmetry too. We are now able to 
proceed in the usual way known in supergravity, and `explain' the maximally
extended supersymmetry as a simple supersymmetry in twelve dimensions,
$$ SO(2,2) \otimes SO(8) \subset SO(2,10)~.$$
Note that the $2+10$ dimensions are the nearest ones in which the Majorana-Weyl
spinors and self-dual tensors also appear, like in $2+2$ dimensions. It should
be noticed that twelve dimensions for string theory were originally motivated 
in a very different way, namely, by a desire to explain the S-duality of type 
IIB string in ten dimensions as the T-duality of the 12-dimensional string 
dimensionally reduced on a two-torus~\cite{hull}. The type IIB string is then
supposed to arise upon double dimensional reduction from the hypothetical 
$2+10$ dimensional F-theory.

\section{M-branes and their symmetries}

Since the full covariant $2+2$ dimensional action describing M-branes is still
unknown, the first step towards its construction is to determine the 
world-volume and target-space symmetries it should possess. Since in the 
light-cone gauge it is supposed to describe the self-dual gauged $N=8$ 
supergravity,~\footnote{One may distinguish between the `closed' and `open' 
M-branes corresponding to the maximal \newline ${~~~~~}$ $N=8$ and $N=4$
 world-volume 
supersymmetry, respectively.} and the latter actually possess the larger 
{\it superconformal} symmetry $SL(4|8)$~\cite{sie}, it should also be the 
fundamental world-volume symmetry for the M-branes. It simply follows from the
facts that the conformal extension of the `Lorentz' group $SO(2,2)$ is given 
by the group $SL(4)\cong SO(3,3)$, while its $N$-supersymmetric extension is 
the superconformal group $SL(4|N)$. Accordingly, one should use the symmetry
$SSL(4|4)$ for the `open' M-branes. The six dimensions which are known to be
distinguished by the string-string duality, are also distinguished for 
describing M-branes since the $2+2$ dimensional superconformal group has a
unique {\it linear} representation only in six dimensions. Since the internal
symmetry of the $N=4$ superconformal group is $SL(4)\cong SO(3,3)$, combining
it with the $2+2$ dimensional conformal group $SO(3,3)$ also implies `hidden' 
twelve dimensions in yet another way: $SO(3,3)\otimes SO(3,3) \subset SO(6,6)$.
The $6+6$ dimensions is the only alternative to $2+10$ dimensions where
Majorana-Weyl spinors also exist.

It should be noticed that we didn't recover the full $SO(2,10)$ symmetry, which
is the natural `Lorentz' symmetry in $2+10$ dimensions, but some of its
natural decompositions. It may be related to the fact that there is no 
covariant supergravity theory in $2+10$ dimensions. Twelve dimensions may
however be useful as a book-keeping device at least. A natural way to deal 
with the non-covariance problem  may be to employ a null 
reduction~\cite{ov,kmo} which effectively reduces the target space of 
M-branes down to $1+10$ dimensions, thus making a contact to the 
11-dimensional supergravity and M-theory. It is also worthy to mention that 
the group $SO(2,10)$ is the conformal group for $1+9$ dimensions.

\section{Conclusion}

Our arguments support the idea~\cite{kmo} that the unifying framework for
describing the M-theory is provided by the $2+2$ dimensional supermembrane 
theory in $2+10$ dimensions. Self-duality of membranes naturally substitutes 
and generalizes the conformal symmetry of string world-sheet. Our basic 
symmetry requirements were merely the linearly realised 'Lorentz' invariance 
and `space-time' supersymmetry in $2+2$ dimensions. A new feature is the
presence of the maximal world-volume conformal supersymmetry. Although this 
conformal symmetry is non-linearly realized in $2+2$ dimensions, there exists 
its linear realization in six dimensions. The target space dimension $(12)$ is
maximal in the sense that it it accommodates all known supergravity theories, 
as well as the maximal number $(8)$ of Majorana-Weyl spinor supercharges. 

The superconformal symmetries of M-branes should be responsible for their full
integrability and the absence of UV divergences in $2+2$ dimensions. Even
though a four-dimensional M-brane action is expected to be non-linear and, 
hence, non-renormalisable as a quantum theory, it may still be UV finite. For
example, the maximally supersymmetric DNS non-linear sigma-model is likely to
be UV finite in $2+2$ dimensions, as was recently argued in ref.~\cite{ke2}. 
The M-brane theory should actually have an even larger underlying symmetry 
given by an {\it affine} extension of the (super)conformal symmetry, which is 
known to be hidden in the DNS theory~\cite{yale}, and in self-dual field 
theory equations as well~\cite{frei}.

Our way of reasoning unifies all strings and superstrings with any 
number of world-sheet supersymmetries towards the M-theory. It seems to be the
good alternative to the `conformal' embeddings proposed earlier~\cite{em}. All
string theories now arise by combining a compactification of the M-brane with 
a GSO projection. 

Unlike the N=2 strings facing severe infra-red problems in loop 
calculations~\cite{gbi}, there are no such problems for M-branes due to the 
higher world-volume dimension and supersymmetry. The theory of M-branes should
therefore exist as a quantum theory, in which strings would appear as 
asymptotic states of M-brane.

\section*{Acknowledgement}

\noindent I would like to thank the Theory Division of CERN in Geneve, where
this work was completed, for hospitality extended to me during the Workshop
on String Duality II, in June 17--21, 1996.

\end{document}